\documentclass[aps,prb,showpacs,twocolumn,superscriptaddress]{revtex4}
\usepackage{graphicx,bm,amsmath,amssymb,psfrag}
\usepackage{color}

\begin{document}
\title{All electron topological insulator in InAs double wells}
\author{Sigurdur I.\ Erlingsson}
\email{sie@ru.is}
\affiliation{School of Science and Engineering, Reykjavik University, Menntavegi 1, IS-101 Reykjavik
  Reykjavik, Iceland}
\author{J.\ Carlos Egues}
\affiliation{Instituto de F\'{\i}sica de S\~ao Carlos, University of S\~{a}o Paulo, 13560-970
  S\~{a}o Carlos, SP, Brazil}
\begin{abstract}
We show that electrons in ordinary III-V semiconductor double wells with an in-plane modulating periodic potential and inter well spin-orbit interaction are tunable Topological Insulators (TIs). Here the essential TI ingredients, namely, band inversion and the opening of an overall  bulk gap in the spectrum arise, respectively, from (i) the combined effect of the double well even-odd state splitting $\Delta_{SAS}$ together with the superlattice potential and (ii) the interband Rashba spin-orbit coupling $\eta$. We corroborate our exact diagonalization results by an analytical nearly-free electron description that allows us to derive an effective Bernevig-Hughes-Zhang (BHZ) model. Interestingly, the gate-tunable mass gap $M$ drives a topological phase transition featuring a discontinuous Chern number at $\Delta_{SAS}\sim 5.4$\, meV. Finally, we explicitly verify the bulk-edge correspondence by considering a strip configuration and determining not only the  bulk bands in the non-topological and topological 
phases but also the edge states and their Dirac-like spectrum in the topological phase. The edge electronic densities exhibit peculiar spatial oscillations as they decay away into the bulk. For concreteness, we present our results for InAs-based wells with realistic  parameters.
\end{abstract}
\maketitle

\section{Introduction}
Topological Insulators (TI) have been theoretically predicted in graphene \cite{kane05:226801} and in negative-gap or inverted-band HgTe-based quantum wells \cite{bernevig06:1757}, being experimentally realized in the latter shortly after \cite{konig07:766}. These are exotic solids being bulk insulators with metallic edges or surfaces \cite{hasan-kane,qi11:1057}.  An essential ingredient for a system to exhibit TI phases
is the existence of a tunable bulk band gap that can not only be tuned to zero but also invert its sign. Defining as ``positive gap'' the case in which it can be mapped without closing onto the $m_0c^2>0$ gap separating the positive- and negative-energy solutions of the Dirac equation (``vacuum''), one can immediately see that an interface between materials with positive and negative gaps must support gapless (edge or surface) states \cite{pankratov87:93}. Time reversal symmetry and spin orbit interaction inextricably lock the spin and momentum of these states making them  
helical \cite{koenig}.

More recently, interesting works have proposed TIs with ordinary bulk materials \cite{liu08:236601,patrik,sushkov13:18601, dong}, as naturally occurring inverted-band or negative-gap materials are usually unconventional narrow band-gap systems. These proposals rely on externally inducing a band inversion of the electron and hole states, e.g., electrically in double well systems \cite{liu08:236601,knez, patrik}.  Hexagonal patterns fabricated in $p$-doped GaAs well to resemble the physics of Dirac carriers in graphene offer
another means to attain TI phases\cite{sushkov13:18601}. Another appealing idea is the use of built-in polarization fields to induce band inversion and TI phases in Ge sandwiched between GaAs layers \cite{dong}. All of these works rely on electrons and holes \cite{liu08:236601,patrik,dong} or holes only \cite{sushkov13:18601}
\begin{figure}[t]
 \begin{center}
   {\vspace{-0.75cm}\includegraphics[angle=0,width=0.22\textwidth]{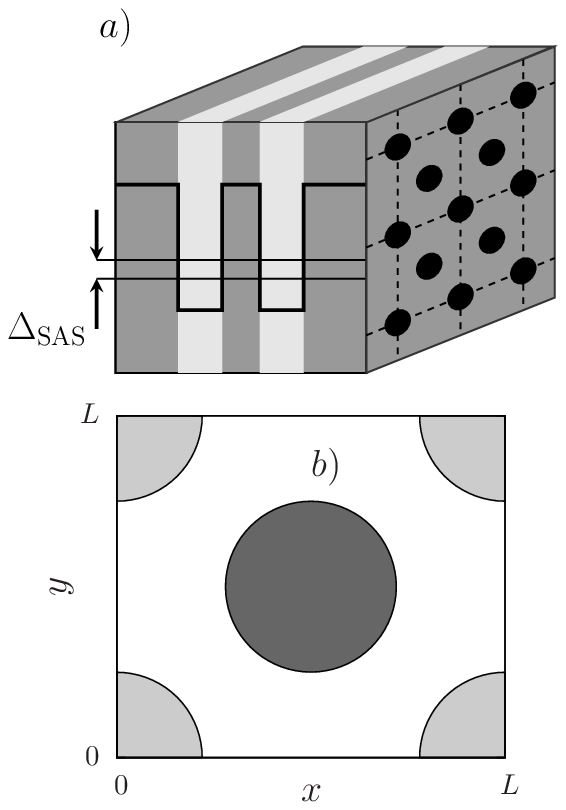}}\!\!\!\!{\vspace{0.25cm}\includegraphics[angle=0,width=0.22\textwidth]{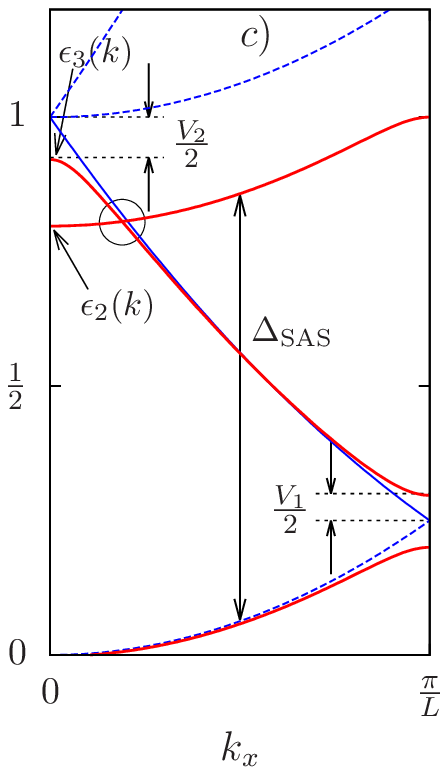}}
  \caption{a) 
  Proposed double well structure  and its potential profile showing the two lowest states split by $\Delta_\mathrm{SAS}$. The in-plane superlattice is denoted by the black dots on top of the structure. b) Unit cell of the periodic ``pits'' in a).  The relative depth of the pits, indicated by different shadings, determine the values $V_1$ and $V_2$ that parametrize the periodic potential \cite{footnoteSymmetry}. c) Free and nearly-free electron bands (dashed and solid, respectively) in units of $\frac{\hbar^2Q^2}{2m^2}$ along the $k_x$ axis. The interband spin-orbit coupling $\eta$ will open up a gap at the crossing (circle) of the inverted bands, see Fig. \ref{fig:GammaXM}.}
  \label{fig:schematic}
  \vspace{-0.5cm}
  \end{center}
\end{figure}

\begin{figure}[t]
\begin{center}
\includegraphics[angle=0,width=0.85\columnwidth]{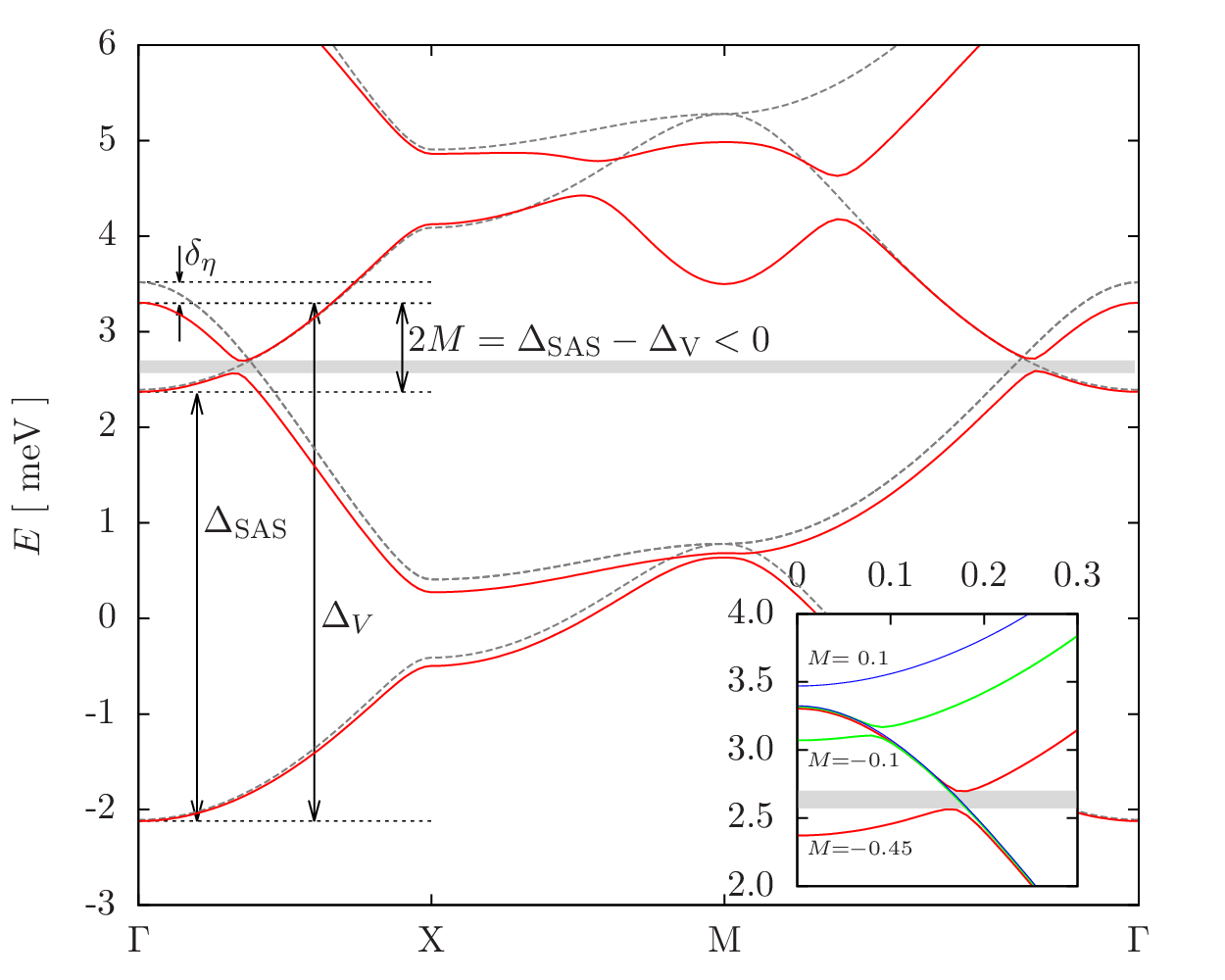}
\caption{In-plane superlattice bandstructure (red curves) for $V_1=3.5$\,meV, $V_2=12$\,meV, $\Delta_{SAS}=4.5$\,meV and $\eta=20$\,meV nm.  The gray (dashed) curves are the energy bands for $\eta=0$. A finite interband spin-orbit coupling $\eta$ gives rise to anti crossing of the inverted bands, thus generating an overall gap (shaded rectangular gray area). The shift $\delta_\eta$ is due to ISOC corrections to the band energies.  The inset shows the bandstrucure for different values of $M (\Delta_\mathrm{SAS})=-0.45 (4.5), -0.1 (5.2)$, and $0.1 (5.6)$\,meV.}
\label{fig:GammaXM}
\end{center}
\end{figure}

Here we propose a TI based on ordinary III-V semiconductor nanostructures with only electrons. We consider 
a bilayer quantum well with two confined electron subbands and intersubband spin-orbit coupling (ISOC) \cite{calsaverini08:155313,she}, Fig.\ \ref{fig:schematic}a. Such even/odd two band systems have been realized in wide quantum well \cite{hernandes}. By fabricating a periodic pattern on top of the structure (e.g.\ via etching) and depositing metal gates  gives rise to  an in-plane  modulating  superlattice potential within the QW \cite{schlosser}, Fig.\ \ref{fig:schematic}a, we are able to obtain the necessary ingredients for a TI: (i) tunable inverted subbands controlled by the `mass gap' $2M = \Delta_{SAS} - \Delta_V$ that depends on both the double-well even-odd state splitting  $\Delta_{SAS}$  and
the quantity $\Delta_V$ that is determined by the gate controllable parameters of the periodic potential, 
and (ii) a bulk overall  gap controlled via the ISOC $\eta$ that gives rise to anti crossings, see bands around the $\Gamma$-point in Fig.\ \ref{fig:GammaXM}. In principle the Fermi energy can be tuned so as to lie in the bulk gap.
The sign of $M$ can be tuned either through $\Delta_\mathrm{SAS}$ (different quantum well structures) or the gate-controllable $\Delta_V$, see Appendix \ref{sec:DeltaSAS}.

We solve the problem within the physically appealing nearly-free electron description. In this approach we analytically derive an effective BHZ model for our system \cite{bernevig06:1757}. 
We also solve the problem numerically via exact diagonalization thus determining the full energy spectrum within the Brillouin zone. In the appropriate parameter range the two descriptions agree very well. 
We also calculate the topological invariant from our bulk band structure and show that the system undergoes a topological phase transition when the mass $M$ changes sign, indicated by a discontinuity in the topological invariant as a function of $\Delta_\mathrm{SAS}$ or $\Delta_V$, see Fig. \ref{fig:dxyzPlot}.
We then find the  solutions for a strip configuration to verify the bulk-edge correspondence explicitly: in the non-topological phase ($M>0$) our system is a bulk insulator with no edge states while in the topological regime 
($M<0$) it features, in addition, gapless edge states with Dirac-like bands, see Fig.\ \ref{fig:edgeDispersionPlot}. Interestingly, we find that the edge states display oscillations as they spatially decay away from the border into bulk, see Fig.\ \ref{fig:edgeDensityPlot}. These oscillations can in principle be mapped via scanning gate microscopy recently used to probe edge states in HgTe-based TI wells \cite{koenig}.

\section{Model system} 
Our effective $4\times 4$ model describes the two lowest subbands of a symmetric quantum well (plus spin). In the basis $\{ |\bm{p},e\uparrow \rangle, |\bm{p},o\downarrow \rangle, |\bm{p},e\downarrow \rangle,|\bm{p},o\uparrow \rangle \}$  of the (e)ven and (o)dd eigenstates we have 
\begin{eqnarray}
 H_w(\bm{p})
 \!\!\!\!&=&\!\!\!\! \left (
\begin{array}{cccc}
\frac{ p^2}{2m^*} & -i\frac{\eta}{\hbar} p_-& 0 & 0 \\
i\frac{\eta}{\hbar} p_+& \frac{ p^2}{2m^*}+\Delta_{SAS} & 0 &0 \\
0 &0 &\frac{ p^2}{2m^*} & i\frac{\eta}{\hbar} p_+ \\
0 &0 & -i\frac{\eta}{\hbar} p_- & \frac{p^2}{2m^*} +\Delta_{SAS}
\end{array}
\right ),
 \label{eq:QW}
\end{eqnarray}
where $\bm{p}$ denotes the electron momentum and $p_\pm=p_x\pm i p_y$.  Here $\eta$ is the interband spin-orbit coupling \cite{calsaverini08:155313} and $m^*$ the electron effective mass. 
The Hamiltonian in (\ref{eq:QW}) can be put into the standard BHZ form by the unitary transformation
$U=e^{-\pi \sigma_y/4}$ acting on each $2\times 2$ diagonal block. This results in the new $2\times 2$ upper diagonal block
\begin{eqnarray}
H_{w,2\times 2}(\bm{p})
&=&
\left (
\begin{array}{cc}
 \frac{p^2}{2m^*}+\Delta_{SAS} & \frac{\eta}{\hbar} p_+  \\
 \frac{\eta}{\hbar} p_- &\frac{p^2}{2m^*} 
\end{array}
\right ) .
\label{eq:HQW2x2}
\end{eqnarray}
The lower diagonal block is the time reversed version of Eq.\ (\ref{eq:HQW2x2}): $H^*_{w,2\times 2}(-\bm{p})$.  Equation (\ref{eq:HQW2x2}) has a form reminiscent of the $2\times2$ diagonal blocks of the BHZ model, apart from the inverted band structure \cite{bernevig06:1757}.  In the BHZ model the inverted band structure arises from the peculiar ordering of bands of HgTe combined with the tunability of the electron and hole levels in a well geometry.  Here we will {\em engineer} an inverted-band system from the ordinary double well with normal ordering of bands by superimposing a two dimensional superlattice on top of it, Fig. \ref{fig:schematic}a and b. We choose the periodic potential (period $L$) as
\begin{eqnarray}
V(\bm{r})=V_1 \left (\cos(Q x)+\cos(Q y) \right ) +V_2\cos(Q x)\cos(Q y),
\label{eq:Vxy}
\end{eqnarray}
where $ Q=\frac{2\pi}{L}$.  
This potential gives rise to parabolic dispersions around the $\Gamma$-point with positive curvature (mass)
for the first and second subbands and negative curvature for the third band, schematically shown in Fig.\ \ref{fig:schematic}c.
More specifically, the superlattice Hamiltonian is
\begin{eqnarray}
H_\mathrm{SL}&=&H_{w,2\times 2}(-i\hbar \partial_x,-i \hbar\partial_y)+V(x,y) \mathbb{I}_{2\times 2} .
\label{eq:Hfull}
\end{eqnarray}
where we have used $\bm{p}=-i \hbar \bm{\nabla}$.  The corresponding eigensolutions are Bloch wave functions  $\psi_{\bm{k},n}(\bm{r})=e^{i\bm{k}\cdot \bm{r}} u_{\bm{k},n}(\bm{r})$ with energies $\varepsilon_{n}(\bm{k})$, $u_{\bm{k},n}(\bm{r})$ has the same periodicity as $V(\bm{r})$ \cite{kittel95:book}. It is convenient to define the energy scale  $E_Q=\frac{\hbar^2Q^2}{2m^*}$, which for InAs ($m^*=0.022$) and, say, superlattice period $L=80$\,nm, yields $E_Q \approx 10$\,meV.

\section{Gapped bulk spectrum: Numerics} 
Figure\ \ref{fig:GammaXM} shows the band structure (red curves) obtained via exact diagonalization using the parameters: $V_1=3.5$\,meV, $V_2=12.0$\,meV,  $\Delta_{SAS}=4.5$\,meV, and $\eta=20$\, meVnm \cite{calsaverini08:155313}. The interband coupling $\eta$ can be further increased by optimizing the quantum well structure \cite{li08:152107}. 

The $V_2$ term opens up a gap at the $\Gamma$-point, giving rise to a negative curvature band, and $V_1$ facilitates the coupling between the second and third states for finite $\bm{k}$ values, see Eqs.\ (\ref{eq:pert1}) and (\ref{eq:pert2}). 
When $V_2 \approx E_Q$ the gap opens up over the full Brillouin zone.
The gray dashed curves show the bands in the absence of the spin-orbit coupling $\eta$.  The 2nd and 3rd bands clearly show inversion and crossings for $\eta=0$, while a non-zero $\eta$ opens up gaps at the crossing (red curves).  
The energy splitting of the inverted bands is given by the tunable mass gap $2M=\Delta_\mathrm{SAS}-\Delta_V$.  The parameter $\Delta_V$ is defined as the energy difference between the 1st and 3rd energy bands at the $\Gamma$-point, see Fig.\ \ref{fig:GammaXM}.
The value of the even-odd energy splitting $\Delta_{SAS}$ is controlled by the structure of the quantum well confining potential.  The inset in Fig. \ref{fig:GammaXM} is a blowup of the band crossing for $\Delta_V=5.4$\,meV and three different values of $\Delta_{SAS}$, going from an inverted ordering of bands for $M=-0.45$\,meV and $-0.1$\,meV to a normal ordering at $M=0.1$\,meV.
The bands in Fig.\ \ref{fig:GammaXM} are doubly degenerate due to the time-reversed part of the full Hamiltonian, i.e.\ the Kramers pairs.

\section{Nearly-free electron description} 
Here we focus on the 2nd and 3rd bands for $\eta=0$ (see gray curves in Fig.\ \ref{fig:GammaXM}), which comprise the two inverted crossing bands required by the BHZ model. To obtain analytical results and a better qualitative understanding of our system we now follow a perturbative approach based on the nearly free electron model (NFEM).

\subsection{Energy bands and controlled band inversion}
We start by looking at the single quantum well in the presence of a periodic potential.  
Using Bloch's theorem, the eigenvalue problem corresponding to the Hamiltonian in Eq.\ (\ref{eq:Hfull}) reduces to
\begin{eqnarray}
& &\left ( \frac{\bm{p}^2}{2m^*}+\frac{\hbar}{m^*}\bm{k}\cdot \bm{p}+V(x,y) \right )u_{n,\bm{k}}(x,y) \nonumber \\
&=&\left ( \varepsilon_{n,\bm{k}}-\frac{\hbar^2k^2}{2m^*} \right )u_{n,\bm{k}}(x,y).
\end{eqnarray}
In the $\bm{k}\cdot \bm{p}$ spirit, we are interested in finding the energy spectrum at the $\Gamma$-point ($\bm{k}=0$) and then use these states to calculate the energy bands away from the $\Gamma$-point, using pertubation theory to obtain corrections due to $\frac{\hbar}{m^*}\bm{k}\cdot \bm{p}$.  
For the free electron model, the energy levels at the $\Gamma$-point occur at $\varepsilon=0$ (1 state), $\varepsilon=E_Q$ (4 degerate states), $\varepsilon=2E_Q$ (4 degerate states), etc.\ , where $E_Q=\hbar^2 Q^2/2m^*$.   
To describe the inverted bands we need to consider the four normalized states that cross at $\varepsilon=E_Q$
\begin{eqnarray}
 \left \{\frac{e^{iQx}}{L},\frac{e^{iQy}}{L},\frac{e^{-iQx}}{L},\frac{e^{-iQy}}{L} \right \}, \label{eq:basis2}
\end{eqnarray}
and the ground state at $\varepsilon_1$, i.e.\ $ \left \{ \frac{1}{L} \right \}$.
The 4 states in Eq.\ (\ref{eq:basis2}) are coupled by the $V_2$ term 
and using degenerate perturbation theory the new states and corresponding eigenenergies are
\begin{eqnarray}
u_{2,A}(x,y)&=&\frac{1}{2L}(e^{iQx}+e^{-iQx}-e^{iQy}-e^{-iQy}),\label{eq:2A}\\
u_{2,B}(x,y)&=&\frac{1}{\sqrt{2}L}(e^{iQx}-e^{-iQx}), \\
u_{2,C}(x,y)&=&\frac{1}{\sqrt{2}L}(e^{iQy}-e^{-iQy}), \\
u_{2,D}(x,y)&=&\frac{1}{2L}(e^{iQx}+e^{-iQx}+e^{iQy}+e^{-iQy}),
\end{eqnarray}
and the corresponding eigenergies are $\varepsilon_{2,A}=E_Q-V_2/2$, $\varepsilon_{2,B}=E_Q$, $\varepsilon_{2,C}=E_Q$, and $\varepsilon_{2,D}=E_Q+V_2/2$.  The state $u_{2,A}(x,y)$, that gets lowered in energy by $V_2/2$, along with the ground state that we denote by $u_1(x,y)=1/L$ (eigenenergy $\varepsilon_1=0$) will form the inverted bands when the even/odd state energy separation in the bilayer system is considered. 

\begin{figure}[h]
\begin{center}
\includegraphics[angle=0,width=0.45\textwidth]{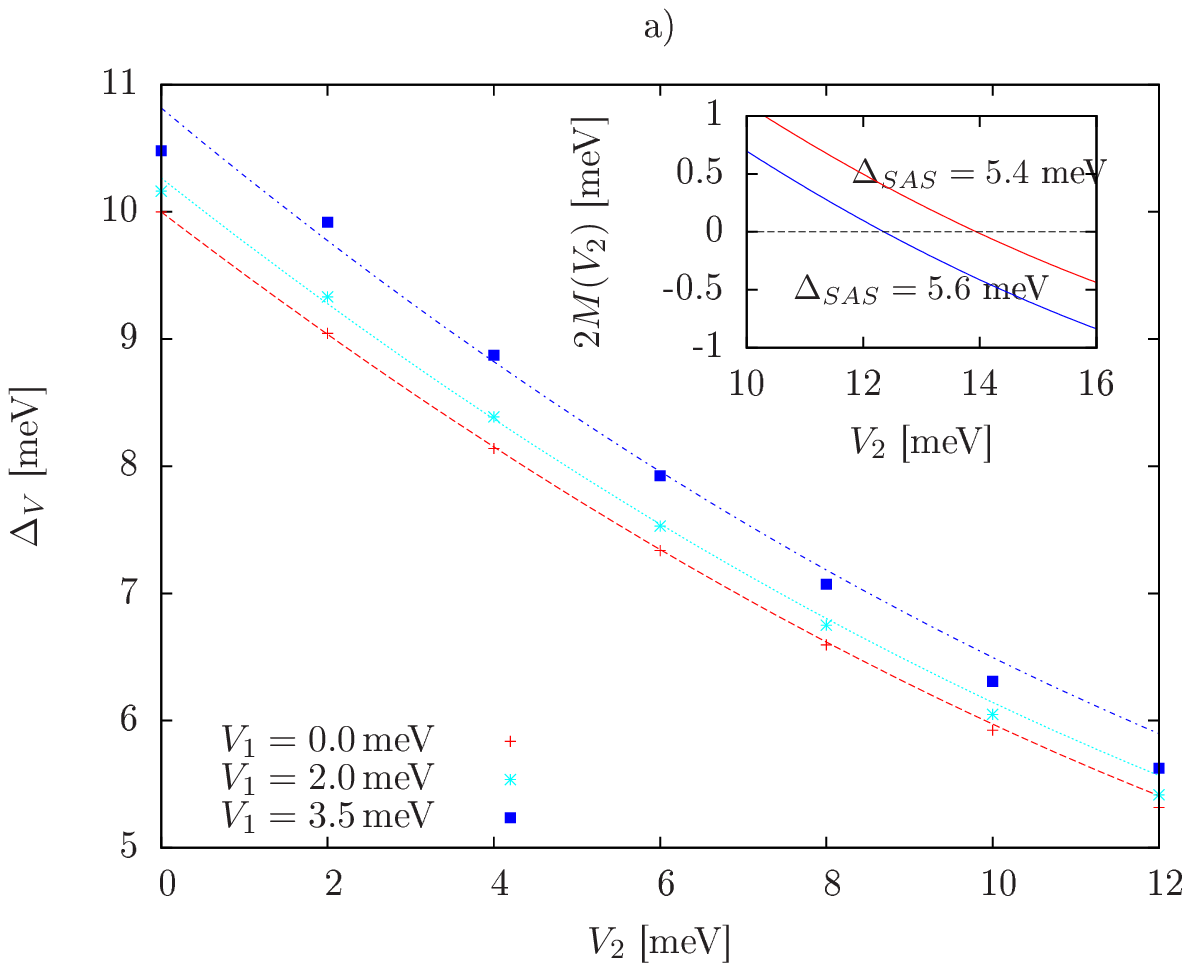}
\includegraphics[angle=0,width=0.45\textwidth]{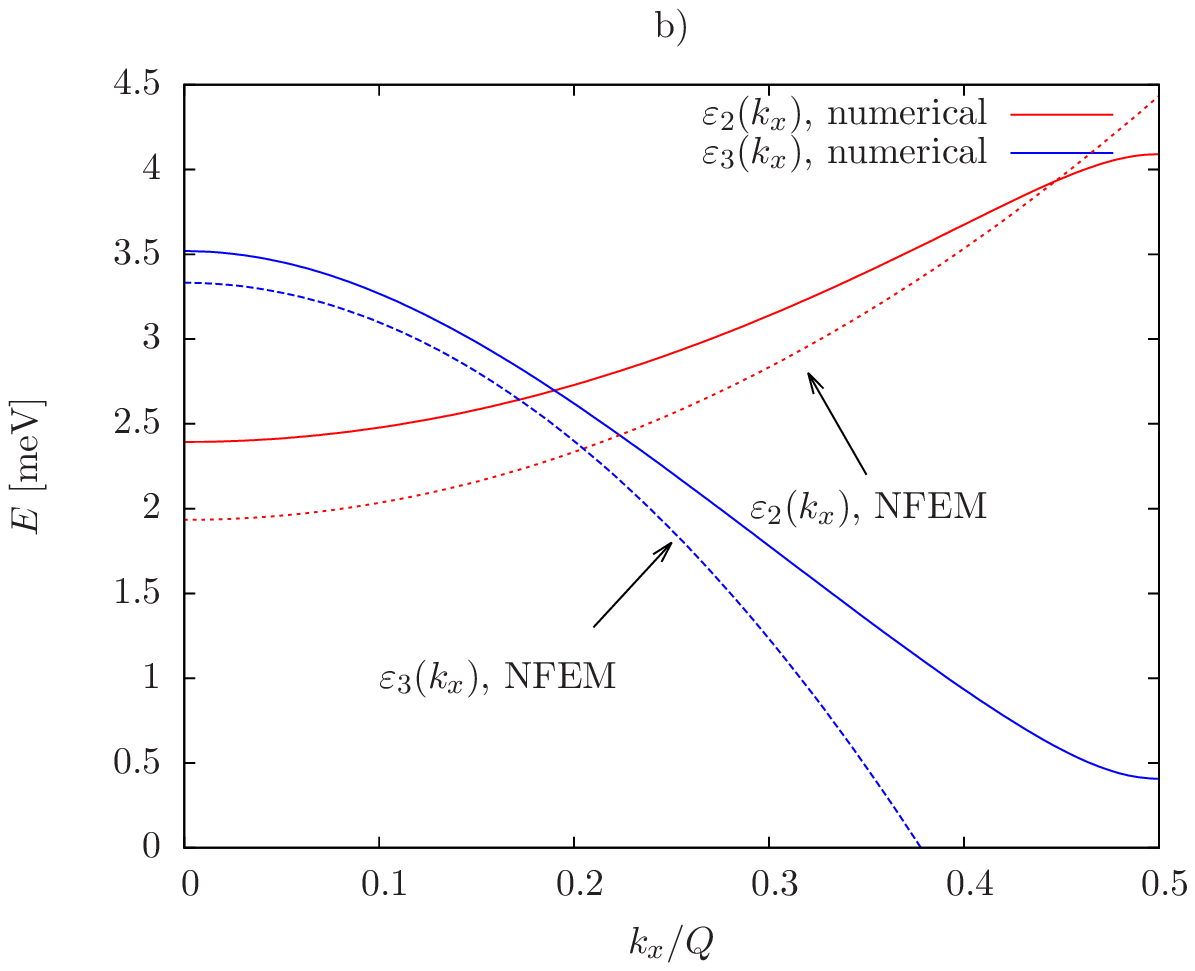}
\caption{a) The bandwidth $\Delta_V$ as a function of $V_2$, for three values of $V_1$.  The dashed curves are the pertubative result obtained using Eqs.\ (\ref{eq:E1G}), (\ref{eq:E3G}) and (\ref{eq:DeltaV}).  The inset shows the mass gap $2M=\Delta_\mathrm{SAS}-\Delta_V(V_2)$ as a function of $V_2$ around the value of $V_2=12$\,meV, showing that the bands can be inverted via the gate that defines the superlattice potential. b) Numerical and pertubative results, Eqs.\ (\ref{eq:E3kx}) and (\ref{eq:E2kx}),  for bands 2 and 3 and $\eta=0$.  A nonzero $\eta$ opens up a gap at the crossing point, see Fig.\ \ref{fig:GammaXM}.}
\label{fig:DeltaV}
\end{center}
\end{figure}

Anticipating the energy spectrum for the bilayer system we simplify the notation and use the same labeling scheme as in Fig.\ \ref{fig:schematic}c) and denote state 2A in Eq.\ (\ref{eq:2A}) by $u_3$ and corresponding eigenenergy $\varepsilon_3$.
The second order perturbation theory correction to states $u_1$ and $u_3$ at $k=0$ are 
%
\begin{eqnarray}
\varepsilon_1(k=0)
&=&0-\frac{V_2^2}{8E_Q}-\frac{V_1^2}{ E_Q+\frac{V_2}{2}} \label{eq:E1G} \\
\varepsilon_3(k=0)
&=&E_Q-\frac{V_2}{2}-\frac{V_2^2}{8}\frac{1}{4E_Q+\frac{V_2}{2}} \nonumber \\
& &-\frac{V_1^2}{4} \left ( \frac{1}{3E_Q+\frac{V_2}{2}} + \frac{1}{E_Q+\frac{V_2}{2}} \right ) \label{eq:E3G}.
\end{eqnarray}
%
The quantity related to the superlattice potential that enters into the mass gap is the bandwidth $\Delta_V$, which is defined as the difference between the top of the second band and the bottom of the first band:
\begin{eqnarray}
 \Delta_V&=&\varepsilon_{3}(k=0)-\varepsilon_{1}(k=0).
 \label{eq:DeltaV}
\end{eqnarray}
%
The NFEM calculation, along with full numerics, of the bandwidth $\Delta_V$ are plotted as a function of $V_2$, for three values of $V_1$ in Figure \ref{fig:DeltaV}a).  In the range corresponding to the values used in the paper the behavior of $\Delta_V$ is predominantly linear in $V_2$ with second order corrections contributing to higher values of $V_1$ and $V_2$.  
An intuitive way, although not mathematically rigorous, to understand why the nearly-free electron model gives qualitatively good results, even for $V_2 > E_Q$, is to write the periodic potential in terms of Fourier components
\begin{eqnarray}
V(x,y)=\frac{V_1}{2}(e^{iQx}+e^{iQy})+\frac{V_2}{4} (e^{iQ(x+y)}+e^{iQ(x-y)} )+\mathrm{c.c.\ }, \nonumber
\end{eqnarray}
which shows that the coupling constants entering the pertubative calculations are effectively $\frac{V_1}{2}$ and $\frac{V_2}{4}$, i.e.\ smaller than $E_Q$, even for $V_2=1.2 E_Q$ as we use in the manuscript.

For simplicity we exhibit the bands along the $k_x$ direction and suppress the $k_y$ varible for clarity. Next we show that the curvature of $\varepsilon_3(k_x)$ is negative.  
Lowest order perturbation in $\frac{\hbar}{m^*}k_x p_x$ results in 
\begin{eqnarray}
\varepsilon_3(k_x)&=&\varepsilon_3(k_x=0)+\frac{\hbar^2k_x^2}{2m^*}+\frac{\left |\frac{\hbar}{m^*} k_x Q \frac{1}{\sqrt{2}} \right |^2}{E_Q-\frac{V_2}{2}-E_Q} \nonumber \\
&=&\varepsilon_3(k_x=0)+E_Q \left (1-\frac{4E_Q}{V_2} \right ) \frac{k_x^2}{Q^2},
\label{eq:E3kx}
\end{eqnarray}
which shows that the curvature is indeed negative for values of $V_2<4E_Q$ \cite{kittel95:book,ashcroft}.  In our calculations we use $V_2=1.2 E_Q$ yielding $\left (1-\frac{4E_Q}{V_2} \right )\approx -2.33$, which compares well to the numerical value of $-2.29$, extracted by fitting the full numerics with a parabolic dispersion.  
The other band that will form the inverted band structure is $\varepsilon_2(k_x)$, which is the same as $\varepsilon_1(k_x)$ apart from a shift in energy of $\Delta_\mathrm{SAS}$.
The curvature of $\varepsilon_2(k_x)$ is simply determined by the free electron dispersion since the $\bm{k}\cdot \bm{p}$ term does not couple the lowest band to any higher-lying bands, resulting in
\begin{eqnarray}
\varepsilon_2(k_x)&=&\Delta_\mathrm{SAS}+\varepsilon_1(k_x=0)+\frac{\hbar^2k_x^2}{2m^*}.
\label{eq:E2kx}
\end{eqnarray}
The two relevant inverted energy bands are shown in Figure \ref{fig:DeltaV}b). 
The NFEM result and the numerics show qualitative agreement, i.e.\ an inverted band ordering and negative curvature of the hole-like band.

When the bilayer quantum well is added together with the superlattice spectrum we get a spectrum similar to the one shown in Fig.\ \ref{fig:schematic}c), i.e.\ the lowest superlattice band of the odd quantum well state is shifted by $\Delta_\mathrm{SAS}$, leading to the band $\varepsilon_2$ in Eq.\ (\ref{eq:E2kx}).  The interband spin-orbit coupling $\eta$ opens up a gap at the crossing point, as shown in the inset of Figure 2.  The size of the anti-crossing gap\cite{qi11:1057} in terms of the BHZ parameters (see App.\ \ref{app:BHZ}) is given by $\Delta_\eta \equiv 2A\sqrt{\frac{M}{B}}$, or term of NFEM parameters
\begin{eqnarray}
\Delta_\eta & \approx &\frac{2\eta}{L} \left [\sqrt{2}\pi\frac{V_1}{V_2}\sqrt{\frac{V_2}{8E_Q^2}(\Delta_V-\Delta_{SAS})} \right ] \nonumber \\
&\approx& 0.1 -0.2\, \mbox{meV}. 
\end{eqnarray}
The size of the gap is predominantly determined by the ratio $\frac{\eta}{L}$ since the quantity in the square brackets is typically of order one. The the quantity in the square brackets is $\approx 0.3$ but that can be more than doubled by optimizing the superlattice parameters.

In addition to opening up the anti-crossing gap, the interband spin-orbit coupling changes slighly the bandwidth $\Delta_V$ since $\varepsilon_3(k_x=0)$ aquires a spin-orbit correction.  This shift can be estimated using second order pertubation theory in $\eta (p_x \pm ip_y)$.  Denoting the bandwidth in the presence of spin-orbit coupling by $\Delta_{V,\eta}$, we can define the spin-orbit induced change in bandwidth as
\begin{eqnarray}
\delta_\eta \equiv \Delta_{V}- \Delta_{V,\eta}=\frac{\eta^2 Q^2}{\Delta_\mathrm{SAS}+\frac{V_2}{2}},
\end{eqnarray}
which gives a calculated spin-orbit induced shift of around $\delta_\eta=\Delta_{V,\eta}-\Delta_V \approx0.23$\,meV for the parameters used in the paper ($\eta=20$\,meV\,nm, $L=80$\,nm, $\Delta_\mathrm{SAS}=4.5$\,meV and $V_2=12$\,meV).  This corresponds quite well to the numerically obtained value of $\delta_\eta=\Delta_{V,\eta}-\Delta_V\approx 0.19$\,meV, see Fig.\ \ref{fig:GammaXM}.

\subsection{Wavefunction away from the $\Gamma$-point} 
The two inverted bands are formed by $\varepsilon_2$ and $\varepsilon_3$ are given in Eqs.\ (\ref{eq:E2kx}) and (\ref{eq:E3kx}).  Due to the square symmetry of the periodic potential the $k_x$ dependence in the bands is the same as for $k_y$, which is also true for the wavefunctions.  The wavefunctions corresponding to $\varepsilon_2$ and $\varepsilon_3$ (for $\eta=0$)  can be found by doing lowest order perturbation theory in $V_1$, $V_2$, and $k_{x,y}$  (see solid lines in Fig.\ \ref{fig:schematic}c)
\begin{eqnarray}
u_{\bm{k},2}(\bm{r})
&=& 1-\frac{2m^*V_1}{\hbar^2 Q^2} (\cos(Qx)+\cos(Qy)) 
\label{eq:pert1} \\
u_{\bm{k},3}(\bm{r})
&=& \cos(Qx)-\cos(Qy)\nonumber \\
 & &+\frac{i\hbar^2 Q}{\sqrt{2} m V_2} \Bigl (k_x \sin(Qx)+k_y \sin(Qy) \Bigr ) .
\label{eq:pert2}
\end{eqnarray}
The zeroth order form of $u_{\bm{k},3}(\bm{r})$ is a linear combination of $\cos(Qx)$ and $\cos(Qy)$ since $V_2$ splits the four degenerate free-electron bands at the $\Gamma$-point.  

\subsection{Effective BHZ model} 
\begin{center}
\begin{figure}[t]
\includegraphics[angle=0,width=\columnwidth]{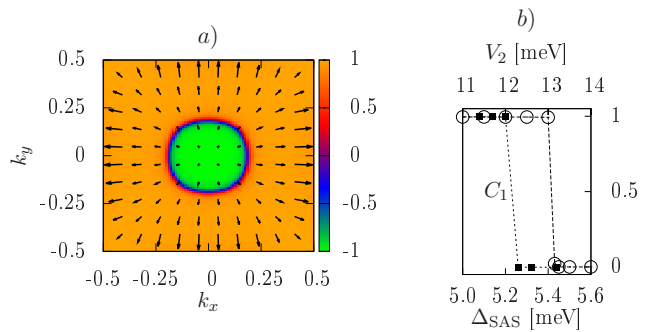}
\caption{Vector field plot of $\bm{d}$ (a) and the topological index $C_1$ (b). In a) the arrows denote the $xy$ components and the color scale the $z$ component of $\bm{d}$, respectively. Note that $d_{x}$ ($d_{y}$) is linear in $k_{x}$ ($k_{y}$) only close to the $\Gamma$-point, in contrast to the BHZ model for which $d_{x}$ ($d_{y}$) is linear irrespective of the value of $\bm{k}$. 
$C_1$ shows a jump as $\Delta_\mathrm{SAS}$ is varied (for fixed $V_1=3.5$\,meV and $V_2=12$\, meV) and when $V_2$ is varied (for a fixed value of $\Delta_{SAS}=5.4$\,meV) }
\label{fig:dxyzPlot}
\end{figure}
\end{center}

We now use the $\eta=0$ states $u_{\bm{k},2}(\bm{r})$, $u_{\bm{k},3}(\bm{r})$ to construct an effective Hamiltonian for $\eta \neq 0$ by projecting the off-diagonal part of Eq.\ (\ref{eq:Hfull}) onto this subspace. These off-diagonal terms are the components of $\bm{d}$-vector of the BHZ model \cite{bernevig06:1757} given by
\begin{eqnarray}
d_{x,y}(\bm{k})
&=&-i\eta \int d^2\bm{r} u_{\bm{k},3}^*(\bm{r})\partial_{x,y}  u_{\bm{k},2}(\bm{r}) 
\label{eq:dxy} 
\end{eqnarray}
The approximate solutions in Eq.\ (\ref{eq:pert1}) and (\ref{eq:pert2}) result in
\begin{eqnarray}
d_{x,y}(\bm{k})
&\approx & A k_{x,y},\quad A \equiv \eta \frac{V_1}{\sqrt{2}V_2}.
\label{eq:dxyPert}
\end{eqnarray}
The properties of the BHZ model, along with the relevant parameters $M$, $A$, etc.\ , are discussed in detail in Refs.\ \onlinecite{bernevig06:1757} and \onlinecite{qi11:1057}, and summarized in App.\ \ref{app:BHZ}. The $z$-component is defined as $d_z(\bm{k})=(\varepsilon_3(\bm{k})-\varepsilon_2(\bm{k}) )/2$,  which corresponds to the separation of $\varepsilon_3(\bm{k})$ and $\varepsilon_2(\bm{k})$ Fig.\ \ref{fig:schematic}c.  In Fig. \ref{fig:dxyzPlot}a) we plot $d_z(\bm{k})/|\bm{d}(\bm{k})|$ (color plot) and $d_x(\bm{k}),d_y(\bm{k})$ (arrows) using the {\it exact} solutions for the energy bands 2 and 3 in Fig.\ \ref{fig:schematic}c) and \ref{fig:GammaXM}.
\subsection{Topological index} 
From the $\bm{d}$-vector we calculate numerically the topological invariant \cite{bernevig06:1757}
\begin{eqnarray}
C_1=\frac{1}{4\pi}\int d^2\bm{k} \hat{\bm{d}}(\bm{k}) \cdot (\partial_1 \hat{\bm{d}}(\bm{k})\times \partial_2 \hat{\bm{d}}(\bm{k}));
\,\hat{\bm{d}}=\frac{\bm{d}}{|\bm{d}|}
\end{eqnarray} 
for $(i)$ fixed superlattice parameters $V_1=3.5$\,meV and $V_2=12$\, meV and varying $\Delta_\mathrm{SAS}$ and $(ii)$ fixed $\Delta_\mathrm{SAS}=5.4$\,meV and varying the (gate-controllable) superlattice parameter $V_2$, keeping $V_1=3.5$\,meV for simplicity. The invariant $C_1$ shows a clear jump when $M$ changes sign, either by varying $\Delta_\mathrm{SAS}$ or $\Delta_V$ via $V_2$, as can be seen in Fig.\ \ref{fig:dxyzPlot}b).

\section{Strip configuration: edge states.} 
Here we verify the bulk-edge correspondence by explicitly finding the edge states of the system in a finite geometry.  We also determine the bulk and edge spectrum for both the topological and non-topological phases, see Fig.\ \ref{fig:edgeDispersionPlot}.
We solve the Hamiltonian in Eq.\ (\ref{eq:Hfull}) for a strip of width $\mathcal{L}_x$, using hard wall boundary conditions $\psi(0,y)=\psi(\mathcal{L}_x,y)=0$.  
Bloch's theorem still applies in the longitudinal $y$-direction. We expand the transverse part of the wavefunction in a normalized sine basis. The number of transverse states is truncated at $M_\mathrm{max}=5 N_\mathrm{per}$ where $N_\mathrm{per}$ is the number of superlattice periods that fit within a strip width.  This corresponds to including, roughly,  $5$ $Q$-vectors in the $x$-direction in the bulk model.  Solving Eq.\ (\ref{eq:Hfull}) for a given value of $k_y$ yields $2\times 5 N_\mathrm{per}$ eigenvalues. Focusing on the eigenvalues in the energy interval corresponding to the BHZ bands, one can plot the relevant set of eigenvalues as function of $k_y$.

\subsection{Gapless edge dispersion }
\begin{center}
\begin{figure}[t]
\includegraphics[angle=0,width=1.1\columnwidth]{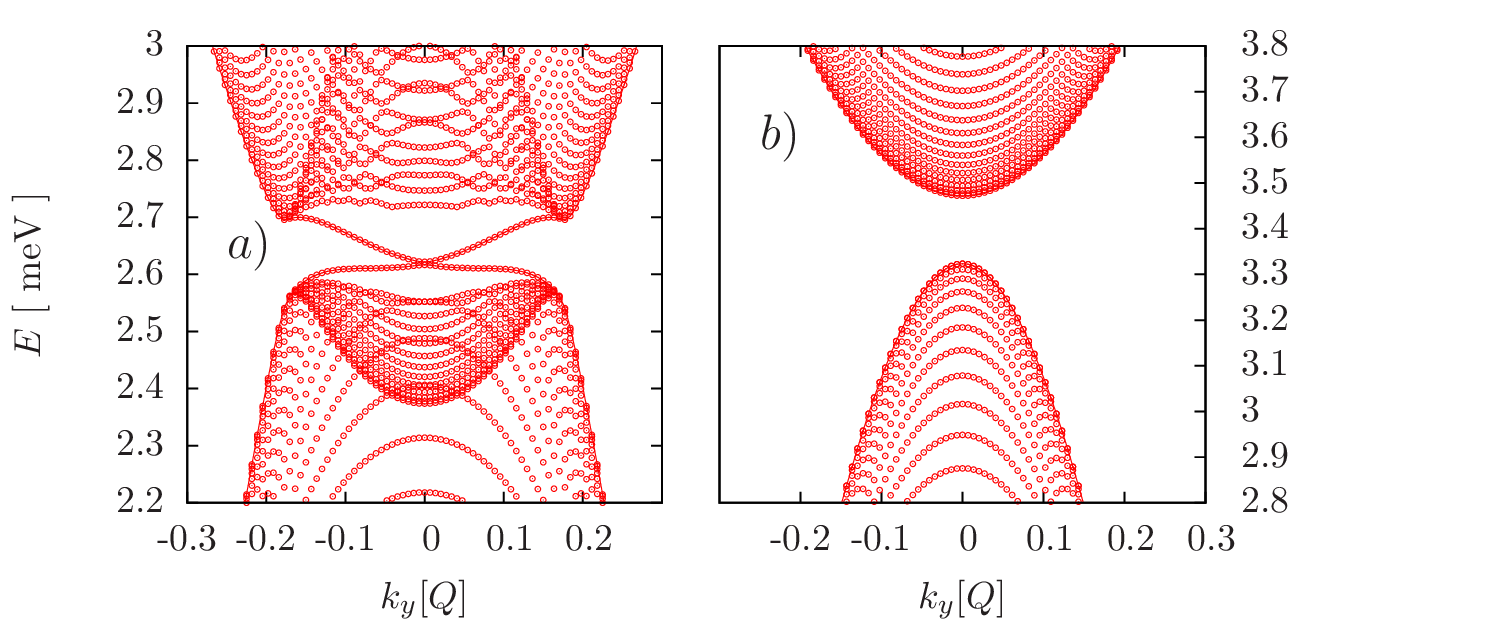}
\caption{Energy spectra resulting from the exact diagonalization of Eq.\ \ref{eq:Hfull}.  a) $\Delta_{SAS}=4.5$\,meV corresponds to the topological phase and edge states are visible while in b) $\Delta_{SAS}=5.6$\,meV gives rise to a normal insulator with no edge states.}
\label{fig:edgeDispersionPlot}
\end{figure}
\end{center}

In Fig.\  \ref{fig:edgeDispersionPlot}a) the value of $\Delta_{SAS}=4.5$\,meV corresponds to the topological phase (see inset in Fig.\ \ref{fig:GammaXM}) and indeed we see edge states in the gap.  In Fig.\ \ref{fig:edgeDispersionPlot}b) the even-odd splitting is increased to $\Delta_{SAS}=5.6$\,meV and the edge states are absent, since the system has now turned to a normal insulator. As can be seen in Fig.\ \ref{fig:GammaXM} $\Delta_{SAS}$ controls the magnitude and the sign of the gap.  Using the BHZ notation, 
the edge state in Fig. \ref{fig:edgeDispersionPlot}a) corresponds to a negative gap $M=d_z(\bm{0})<0$, Fig. \ref{fig:edgeDispersionPlot}b) where $M$ is positive and no gap states appear. Note that for $\Delta_{SAS}=4.5$\,meV our system can not be approximated by an effective BHZ model as the bands show substantial non-parabolicity. This does not influence the existence of the edge states in our system.
\subsection{Oscillatory decaying edges.} 
\begin{center}
\begin{figure}[t]
\includegraphics[angle=0,width=8cm]{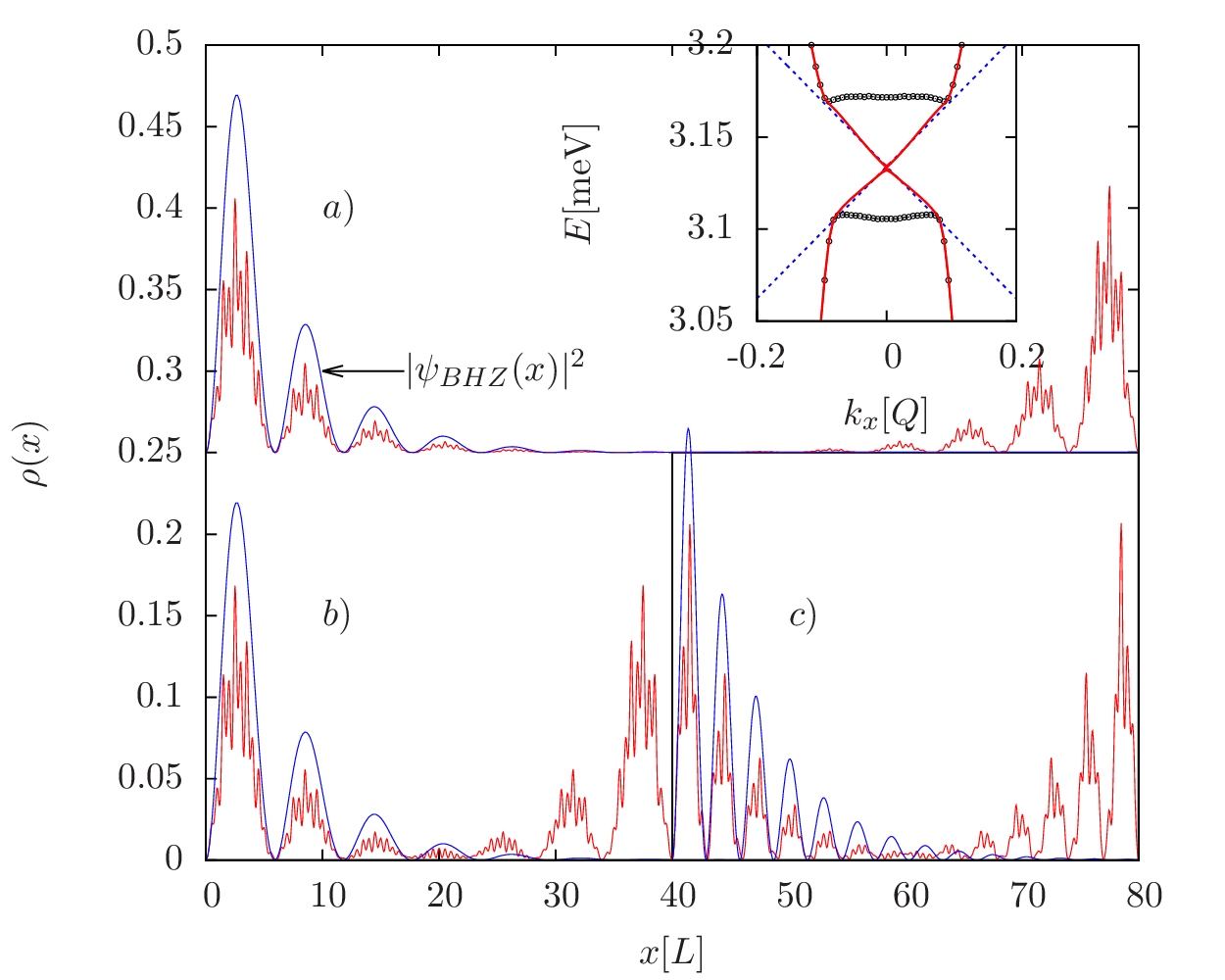}
\caption{
The edge probability density $\rho(x)$ for $\Delta_{SAS}=5.2$\,meV for two strip widths a) $\mathcal{L}_x=80L$ and b) $40L$. The curve for $80 L$ is shifted up by 0.25 for clarity.  The inset in a) shows the edge states along with the BHZ dispersions obtained from the bulk spectrum.  c) Same as in b) but for $\Delta_{SAS}=4.5$\,meV. The blue curves in all panels denote the BHZ results.}
\label{fig:edgeDensityPlot}
\end{figure}
\end{center}

In order to compare our results to the known analytical solution of the BHZ model, we focus on $\Delta_{SAS}=5.2$\,meV.  We can extract the BHZ parameters directly by fitting the curves appearing in the inset of Fig. \ref{fig:GammaXM} to the eigenstates of the BHZ model, i.e.\ Eq.\ (5) in Ref.\ \cite{qi11:1057}.  From these values we can calculate the properties of the edge states using the ansatz $\psi(x) \propto e^{\lambda x}$ \cite{qi11:1057,zhou08:246807,michetti12:124007}.  Solving for $\lambda$ in the middle of the gap ($k_y=0.0$ and $E=-DM/B$) results in
\begin{eqnarray}
\lambda&=&\pm \left ( \frac{A}{2\sqrt{B^2-D^2}}\pm i \sqrt{\frac{M}{B}-\frac{A^2}{4(B^2-D^2)}} \right).
\label{eq:lambda}
\end{eqnarray}
Note that for the parameters extracted from the {\em bulk} spectrum $\frac{M}{B}-\frac{A^2}{4(B^2-D^2)}>0$, which yields a $\lambda$ with a non-zero imaginary part in addition to a real part.  This gives rise to a localized edge state that oscillate spatially. Indeed we see from the numerical diagonalization of Eq.\ (\ref{eq:Hfull}) that the edge state decays into the bulk with  slower oscillations due to the imaginary part of $\lambda$ and rapid oscillations due to the period of the superlattice.  The probability densities of the edge states corresponding to $\Delta_{SAS}=5.2$\,meV for strip widths $\mathcal{L}_x=80 L$ and $\mathcal{L}_x=40 L$ are shown in Figs.\ \ref{fig:edgeDensityPlot}a) and b), respectively.
For  $\Delta_{SAS}=5.2$\,meV the decay length and oscillation wavelength are where $1/\mathrm{Re}\{\lambda\}=11.5L$ and $\frac{2\pi}{\mathrm{Im}\{ \lambda \}}=11.8 L$, respectively.
For a given energy in the gap there are two edge states localized at opposite edges, $\psi_{+ k_y,\uparrow}(x)$ localized around $x=0$ and $\psi_{- k_y,\uparrow}(x)$ localized around $x=\mathcal{L}_x$. The density of the BHZ edge state localized around $x=0$, using parameters extracted from Fig.\ \ref{fig:GammaXM} and $k_y=0$,
\begin{eqnarray}
|\psi_{BHZ}(x)|^2\propto \exp(-2 x \mathrm{Re}\{ \lambda \})\sin^2(\mathrm{Im}\{ \lambda \}x),
\end{eqnarray}
agrees well with the numerical results (blue curves in Fig. \ref{fig:edgeDensityPlot}). Note that the edge state density is symmetric for $k_y=0.0$ \cite{zhou08:246807}.  The helical character of the edge-states comes from the time reversed part of the $4\times 4$ hamiltonian, see discussion below Eq.\ (\ref{eq:HQW2x2}).
Figure \ref{fig:edgeDensityPlot}b) shows the edge state probability density for $\Delta_{SAS}=5.2$\,meV corresponding to $M=-0.1$\,meV, which is around a one fourth of that in Fig.\ \ref{fig:edgeDensityPlot}c) which corresponds to $\Delta_{SAS}=4.5$\,meV.
The density in Fig.\ \ref{fig:edgeDensityPlot}c)  shows an oscillation period close to twice as large, reflecting that oscillations are dominated by the quantity $\sqrt{M/B}$, see Eq.\ (\ref{eq:lambda}). Note that experiments based on detecting the edge transport  \cite{brune12:485} and scanning gate microscopy that can image the modulation of the edge charge density profile \cite{koenig} are feasible.  
A potential drawback of our proposal is the relative small size of the gap $2A\sqrt{M/B} \approx 0.1-0.2$\,meV, see Supplementary Material.  However, what mitigates this is the improved transport properties of high quality In$_x$Ga$_{1-x}$As materials, as compared to HgTe based systems.  Optimizing the quantum wells can yield a 3-fold increase in $\eta$ \cite{Grundler}, tuning the superlattice parameters will give a factor 2, and pushing down the period to $40$\,nm, yields a gap $\sim 2-3$\,meV. This lowers the requirements on the sample mobility and temperature of the experiments. On the positive side, our proposal is unique in that it involves only electrons in standard III-V quantum wells where the parameters leading to band inversion -- the superlattice potential ($V_1$ and $V_2$), Fermi energy, charge density and $\Delta_\mathrm{SAS}$-- are easily controllable.

This work was supported by the Icelandic Research Fund, the Brazilian funding agencies CNPq and FAPESP and PRP/USP within the Research Support Center Initiative (NAP Q-NANO).  SIE would like to acknowledge helpful discussions with H.G.\ Svavarsson and G.\ Thorgilsson for assistance with graphics.

\appendix
\section{The role of $\Delta_\mathrm{SAS}$ and $\Delta_V$}
\label{sec:DeltaSAS}
The mass gap $2M=\Delta_\mathrm{SAS}-\Delta_V$ is controlled by two parameters: $(i)$ the energy splitting $\Delta_\mathrm{SAS}$ of the two lowest even and odd double quantum well states and $(ii)$ the 'bandwidth' of the energy bands $\Delta_V$ introduced in the previous section.

For large enough barriers, which is the case in our double barrier, the width of the central barrier is the major factor in controlling $\Delta_\mathrm{SAS}$.  By varying the barrier thickness the value of $\Delta_\mathrm{SAS}$ can be controlled, but it will be fixed for a given sample.
The even-odd splitting $\Delta_\mathrm{SAS}$ is predominantly determined by the quantum well structure, i.e.\ barrier thickness, quantum well width etc.\ and it is almost unaffected by the presence of the periodic potential. In the absence of the lateral superlattice, and after projecting the full double quantum well Hamiltonian onto the, even/odd subspace results in
\begin{eqnarray}
H_{DQW}&=&-\frac{\hbar^2}{2m}(\partial_x^2+\partial_y^2)\mathbb{I}+\frac{\Delta_\mathrm{SAS}}{2}\tau_z,
\end{eqnarray}
where $\tau_z$ is the Pauli matrix for the double quantum well even-odd subspace.  
We now assume that the electrostatic potential can be written as a periodic function in $x$ and $y$, see Eq.\ (3) in manuscript.  The constants $V_1$ and $V_2$ are replaced by functions $\tilde{V}_1(z)$ and $\tilde{V}_1(z)$
\begin{eqnarray}
V_\mathrm{per}(x,y,z)&=&\tilde{V}_1(z) \left ( \cos(Qx)+\cos(Qy) \right ) \nonumber \\
& &+\tilde{V}_2(z)\cos(Qx)\cos(Qy),
\end{eqnarray}
and when inserted into the Laplace equation $\nabla^2V_\mathrm{per}=0$ \cite{jackson}, results in the following equations
\begin{eqnarray}
\partial_z^2\tilde{V}_1(z)-Q^2\tilde{V}_1(z)=0&,&\tilde{V}_1(0)=V_1, \,\tilde{V}_1(\infty)=0 \\
\partial_z^2\tilde{V}_2(z)-2Q^2\tilde{V}_2(z)=0&,&\tilde{V}_2(0)=V_2, \, \tilde{V}_2(\infty)=0.
\end{eqnarray}
The solution to these equations are
\begin{eqnarray}
\tilde{V}_1(z)&=&V_1 e^{-Qz} \approx V_1 e^{-Qd} (1-Q(z-d)) \label{eq:V1tilde}\\
\tilde{V}_2(z)&=&V_2 e^{-\sqrt{2}Qz} \approx V_2e^{-\sqrt{2}Qd}(1-\sqrt{2}Q(z-d)), \label{eq:V2tilde}
\end{eqnarray}
where $d$ is the distance from the surface to the quantum well.  Here we have also assumed that $Qw\ll 1$, where $w$ is the QW width.  When the Hamiltonian with the full electrostatic potential is projected onto the even-odd substace we get
\begin{eqnarray}
\tilde{H}_\mathrm{SL}&=&-\frac{\hbar^2}{2m}(\partial_x^2+\partial_y^2)+V(x,y)+ \frac{\Delta_\mathrm{SAS}}{2}\tau_z \nonumber \\
&+&\frac{wa_\mathrm{eo}}{L} \left [ V(x,y)+(\sqrt{2}-2) V_2 \cos(Qx)\cos(Qy) \right ]\tau_x, \nonumber \\
\label{eq:DQWper}
\end{eqnarray}
where $V(x,y)$ is given in Eq.\ (\ref{eq:Vxy}) and
\begin{eqnarray}
 a_\mathrm{eo} &=& \frac{2\pi}{w} \int dz \chi^*_e(z) z \chi_o(z),
\end{eqnarray}
is a dimensionless constant of order one coming from the matrix element of Eqs.\ (\ref{eq:V1tilde}) and (\ref{eq:V2tilde}) in the the even and odd basis.  We can disregard the mixing of the even and odd states due to lateral periodic potential  (the $\tau_x$ term) as long as
\begin{eqnarray}
 \frac{w^2}{L^2}\frac{ \mbox{max}\{ V_1^2,V_2^2 \} }{\Delta_\mathrm{SAS}^2} \ll 1. \label{eq:estimate}
\end{eqnarray}
Note that $\Delta_\mathrm{SAS}$ can be controlled by the quantum well structure, e.g.\ it can be made larger by a thinner barrier, so this condition can always be satisfied.  For typical quantum wells the barrier and well thicknesses, are of order $w \approx 10$\,nm and the lower limit of periodic potential is $L\approx 40$\,nm.  So, even for relatively thick barriers and short period superlattice we have $\frac{w^2}{L^2}\sim 0.063$. 
The estimate in Eq.\ (\ref{eq:estimate}) is obtained by comparing the two terms that multiply $\tau_x$ and $\tau_z$ in Eq.\ (\ref{eq:DQWper}).  Since the contribution of the different Pauli matrices are added as squares, the condition for discarding the $\tau_x$ comes from comparing the squares of the two contributions, which leads to Eq.\ (\ref{eq:estimate}).

\section{Connection to BHZ model}
\label{app:BHZ}
In the BHZ model\cite{bernevig06:1757,qi11:1057} the two bands that comprise the inverted bands are written as
\begin{eqnarray}
\varepsilon_{\mathrm{e},\uparrow}(\bm{k})&=&C+M+(D-B)k^2\\
\varepsilon_{\mathrm{h},\downarrow}(\bm{k})&=&C-M+(D+B)k^2.
\end{eqnarray}
where $k^2=k_x^2+k_y^2$. The bands will only cross when $M>0$ and $B>D>0$ or when $M<0$ and $B<D<0$.  The parameter $C$ is simply a trivial energy shift whose value has not impact on the physical properties of the system.  These two bands are then coupled, with a coupling strength $A$, resulting in a $2 \times 2$ matrix
\begin{eqnarray}
H_\mathrm{BHZ}&=&\left (
\begin{array}{cc}
M+(D-B)k^2 & Ak_+  \\
 A k_- &-M+(D+B)k^2
\end{array}
\right )  \\
&=&-Dk^2\mathbb{I}+\bm{d}(\bm{k})\cdot \bm{\sigma},
\label{eq:BHZ2x2}
\end{eqnarray}
where $k_\pm=k_x+i k_y$ and the vector $\bm{d}$ is given by 
\begin{eqnarray}
d_x(\bm{k})&=&A k_x, \\
d_y(\bm{k})&=&A k_y, \\
d_z(\bm{k})&=&M-B k^2.
\label{eq:dVector}
\end{eqnarray}
Here we put the trivial constant $C=0$.  The energy difference between the two bands, in the absence of the spin-orbit coupling, is given by $2 d_z(\bm{k})$.
Note that the full $4\times 4$ full BHZ model is obtained by constructing a block diagonal matrix with the upper diagonal is the above matrix and the lower one the time reversed version of Eq.\ (\ref{eq:BHZ2x2}).
The BHZ parameter $A$ was introduced in Eq. (8) in the manuscript and the other parameters can be related to band parameters in our proposal as follows:
\begin{eqnarray}
 2M&\equiv& \Delta_\mathrm{SAS}-\Delta_V \\
  B&=&-\frac{E_Q}{Q^2} \frac{2E_Q}{V_2} <0 \\
  D&=&\frac{E_Q}{Q^2} \left (1- \frac{2E_Q}{V_2} \right )<0.  
\end{eqnarray}


\begin{thebibliography}{15}
\bibitem{footnoteSymmetry} Different magnitudes of $V_1$ and $V_2$ in Eq.\ (\ref{eq:Vxy}) are achieved by making the central holes deeper or shallower than those at the unit cell corners. This pattern is then inherited by the metal deposited on top thus leading to a modulated electric  potential within the QW upon the application of a proper gate potential.
\bibitem{footnote3}The BHZ parameter values corresponding to $\Delta_{SAS}=5.2$ are $C=3.2$\, meV, $D=1.3\times 10^3$\, meVnm$^2$, $M=0.1$\, meV, $B=2.7\times 10^3$\, meVnm$^2$ and $A=5.2$\, meVnm.
\end{thebibliography}
\end{document}